\documentclass{jpsj3} 
\usepackage{txfonts}
\usepackage{bm}
\usepackage{graphicx}

\newcommand{\Tc}{T_{\mbox{\scriptsize c}}}
\newcommand{\Tca}{T_{\mbox{\scriptsize c1}}}
\newcommand{\Tcb}{T_{\mbox{\scriptsize c2}}}
\newcommand{\Ts}{T_{\mbox{\scriptsize s}}}
\newcommand{\Jc}{J_{\mbox{\scriptsize c}}}

\newcommand{\Cph}{C_{\mbox{\scriptsize ph}}}

\title{Size Dependence of Oxygen-annealing Effects on Superconductivity of Fe$_{1+y}$Te$_{1-x}$S$_{x}$}

\author{Teruo Yamazaki, Tatsuya Sakurai, and Hiroshi Yaguchi}
\inst{Department of Physics, Faculty of Science and Technology, Tokyo University of Science, Noda, Chiba 278-8510, Japan} 

\abst{For the Fe-based superconductor Fe$_{1+y}$Te$_{1-x}$S$_{x}$, superconductivity is induced by annealing treatment in oxygen atmosphere, whereas as-grown samples do not show superconductivity.
We have investigated the sample-size dependence of O$_{2}$-annealing effects in Fe$_{1.01}$Te$_{0.91}$S$_{0.09}$.
The annealing conditions are fixed to be 1 atm,  200\,$^\circ$C, and 2\,hours.
We have carried out magnetic susceptibility and specific heat measurements in order to evaluate the superconducting volume fraction. 
We have found that Fe$_{1+y}$Te$_{1-x}$S$_{x}$ has an optimal size for the induction of bulk superconductivity by O$_{2}$ annealing.
Our results indicate that O$_{2}$ annealing is probably effective near the surface of samples over a length of a few tens of micro meters.
}

\begin{document}

\maketitle

\section{Introduction}

Numerous iron-based superconductors have been reported since the discovery of superconductivity in LaOFeP and La(O$_{1-x}$F$_{x}$)FeAs \cite{kamihara2006iron, kamihara2008iron}.
Among them, iron chalcogenide materials such as FeSe, Fe$_{1+y}$Te$_{1-x}$Se$_{x}$, and Fe$_{1+y}$Te$_{1-x}$S$_{x}$ are categorized as 11-systems\cite{deguchi2012physics}. 
The 11-system attracts researchers' attention because of the simplest crystal structure.
The crystal structure of Fe$_{1+y}$Te is tetragonal and the space group is $P4/nmm$.
Fe$_{1+y}$Te ($y \leq 0.12$) does not show superconductivity, 
but exhibits a structural phase transition to 
a monoclinic ($P2_{1}/m$) phase 
at $\Ts\sim65$ K.  
The transition is accompanied by a magnetic phase transition to a commensurate antiferromagnetic phase with a wave vector of ${\bm q}  = \left( \frac{1}{2}, 0, \frac{1}{2} \right) $.
In the higher iron concentration range of $y \geq 0.12$,  
tetragonal Fe$_{1+y}$Te undergoes a structural phase transition 
to an orthorhombic ($Pmmn$) phase 
 and a magnetic phase transition with an incommensurate magnetic structure with ${\bm q}  = \left(\pm \delta, 0, \frac{1}{2} \right)$, where $\delta=0.38$ for Fe$_{1.141}$Te  \cite{li2009first, bao2009tunable,  hu2009superconductivity, zajdel2010phase, koz2013low}. 
The substitution of S for Te in Fe$_{1+y}$Te suppresses the structural and 
the magnetic phase transition temperatures.
While as-grown samples of Fe$_{1+y}$Te$_{1-x}$S$_{x}$ exhibits no superconductivity 
for all the compositions of $x$ and $y$\cite{mizuguchi2011single},
 superconductivity can be induced in Fe$_{1+y}$Te$_{1-x}$S$_{x}$ by several kinds of treatments.
Examples of such treatments are leaving in the air for a long time\cite{mizuguchi2010moisture}, 
 annealing in O$_{2}$ atmosphere (O$_2$ annealing)\cite{mizuguchi2010evolution, kawasaki2011pressure,awana2011superconductivity}, annealing in S atmosphere \cite{deguchi2014excess},
  and  soaking in hot alcoholic beverages or aqueous organic solutions, e.g. red wine or malic acid\cite{deguchi2011alcoholic,deguchi2012clarification}.
After the soaking treatment in the aqueous organic solutions, an Fe component has been detected in the solutions\cite{deguchi2012clarification}.
Hence, it is inferred that the deintercalation of excess Fe may be a key for the induction of superconductivity in Fe$_{1+y}$Te$_{1-x}$S$_{x}$.

In another 11 system, Fe$_{1+y}$Te$_{1-x}$Se$_{x}$, superconductivity can be also induced by various treatments \cite{noji2012specific, imaizumi2012superconductivity, kawasaki2012phase, sun2012effects, rodriguez2011chemical, deguchi2014excess}. 
Among the treatments, annealing in low-pressure oxygen atmosphere or in vacuum can induce bulk superconductivity, and a large critical current density $\Jc$ and a discontinuity of the specific heat have been observed \cite{sun2012effects, noji2012specific, imaizumi2012superconductivity}.  
By contrast, 
 no evidence for bulk superconductivity in Fe$_{1+y}$Te$_{1-x}$S$_{x}$ has been observed thus far,
  although zero electrical resistivity and large diamagnetic susceptibility in zero-field-cooled (ZFC) condition have been observed in Fe$_{1+y}$Te$_{1-x}$S$_{x}$ samples after similar treatments.
Therefore, it is speculated that the superconducting state could be realized only within particular regions of the samples.

In the present work, we have investigated the sample-size dependence of the O$_{2}$-annealing effects
  by means of ZFC and field-cooled (FC) susceptibility and specific heat measurements.
We have chosen the O$_2$-annealing condition to be 1\,atm, 2\,hours, and 200\,$^\circ$C, following Ref. \citen{mizuguchi2010evolution}.
We observed a jump of the specific heat around the superconducting transition temperature $\Tc$ in relatively small O$_2$-annealed samples.
 To the best of our knowledge, this is the first observation of bulk superconductivity in Fe$_{1+y}$Te$_{1-x}$S$_{x}$.
We deduce that the superconducting region induced by O$_2$ annealing is probably near the surface of samples over a length of a few tens of micro meters.

\section{Experimental Procedure}
 The single crystalline samples of Fe$_{1+y}$Te$_{1-x}$S$_{x}$ used in the present study 
 were prepared by a self flux method described in Ref. \citen{mizuguchi2011single}. 
Fe shot (5N), Te shot (6N), and S shot (6N)  
with a nominal composition of $x=0.2$ were
sealed in evacuated quartz ampoule under atmosphere of 0.3-atm argon.
The ampoule was heated by an electric furnace to 1050\,$^\circ$C.
Subsequently, the temperature was kept for 20 hours 
and then cooled to 650\,$^\circ$C at a rate of $-4\,^\circ$C / h.
All the samples used in this work were obtained from the same batch.
The dimensions of the single crystalline sample for measurements 
were 3\,mm $\times$ 3\,mm $\times$ 0.5\,mm$^3$ (class \#1).
We prepared powder samples by crushing single crystals, and classified them by size  
using sieves into the following dimensions;
 106--250$\,\mu$m (class \#2), 75--106$\,\mu$m (class \#3), 20--75$\,\mu$m (class \#4), 
and 0--20$\,\mu$m (class \#5).
The samples were annealed in O$_{2}$ atmosphere .
The pressure,  temperature,  and duration for the O$_{2}$ annealing 
were fixed to be  1 atm,  200\,$^\circ$C, and 2\,hours, respectively.

The actual composition of the samples were determined to be 
Fe$_{1.01}$Te$_{0.91}$S$_{0.09}$ 
using an electron probe micro analyzer EPMA (JEOL, JXA-8100).
X-ray powder diffraction measurements with Fe-K$_{\alpha}$ radiation were 
performed using RIGAKU, RAD-C.

We measured the magnetic susceptibility using a superconducting quantum interference device (SQUID) magnetometer (Quantum Design, MPMS).
For the magnetic susceptibility measurements, the applied field was 20\,Oe.
The single crystalline sample (sample class \#1) was mounted in a plastic straw so that the applied field is along the longest dimension.
The powder samples (sample class \#2--\#5) were wrapped in cling film and mounted in plastic straws.
We have also measured the specific heat by a thermal relaxation method (Quantum Design, PPMS). 
For specific heat measurements for class \#3 (75--106\,$\mu$m), 
the sample was wrapped in copper foil; the heat capacity and the specific heat of the samples were evaluated by subtracting the heat capacity of the copper foil as described in Ref. \citen{shi2010accurate}.
For the subtraction, the data of the specific heat measured for copper in Ref. \citen{martin1960specific} were used.
For the other samples, pellets, 
 prepared by pressing small pieces of crystals, were used for specific heat measurements.
The samples used for magnetic susceptibility, specific heat, and X-ray diffraction measurements are labeled with a, b, and c, respectively, preceded by the class number, e. g. sample 1a.

\section{Experimental Results}
\begin{figure}[t]
\includegraphics[width=78mm]{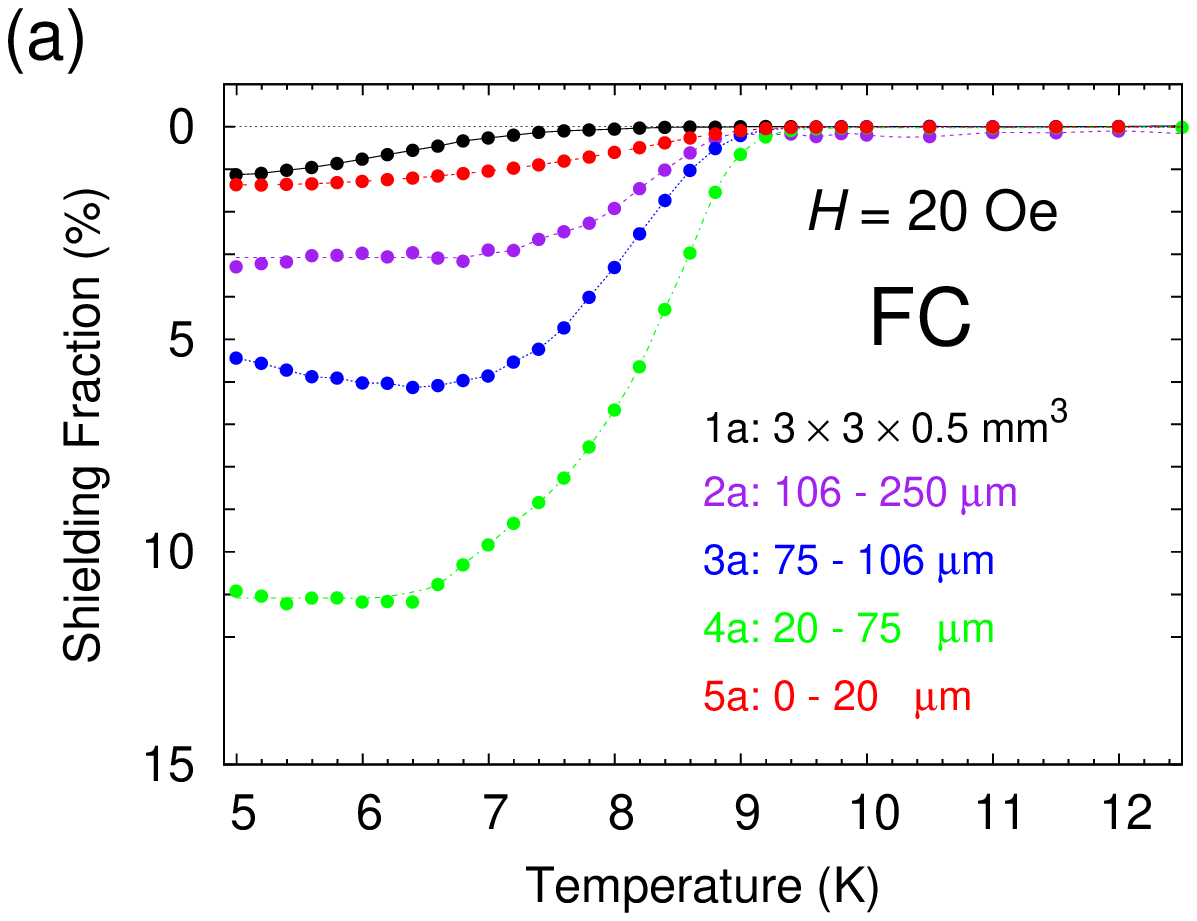}
\includegraphics[width=78mm]{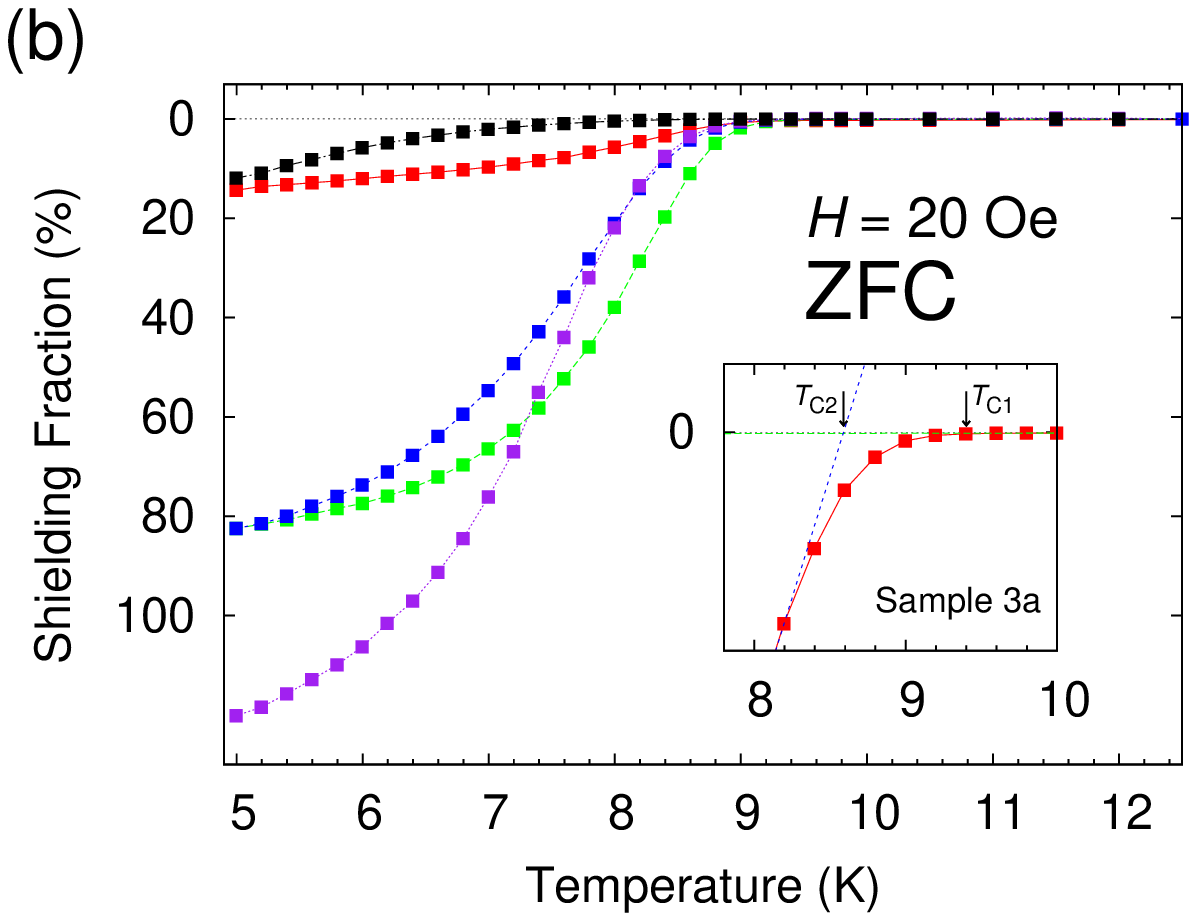}
\caption{(Color online) Temperature dependences of shielding fraction (SF) in O$_{2}$-annealed Fe$_{1.01}$Te$_{0.91}$S$_{0.09}$.
Inset shows an example of the definition of onset $\Tca$ and onset $\Tcb$ for sample 3a. 
Onset $\Tca$ is defined by the start point of the drop of the magnetic susceptibility. 
Onset $\Tcb$ is defined by the intersection of the straight fitting line of the superconducting region with that of the normal state region.
 }
\label{fig:SF-T}
\end{figure}

Figures \ref{fig:SF-T}(a) and (b) show 
the temperature dependences of the superconducting shielding fraction (SF)
of size-classified O$_{2}$-annealed Fe$_{1.01}$Te$_{0.91}$S$_{0.09}$
 evaluated from the  magnetic susceptibility, $\Delta \chi(T)=\chi(T)-\chi(15$\,K),  measured 
 (a) in the zero-field-cooled (ZFC) condition and (b) field-cooled (FC) condition.
In this work, the demagnetizing field effects are ignored.
The onset of the superconducting transition temperature  
 (onset $\Tca$) for single crystalline sample 1a is 8.8 K, 
and the onset $\Tca$ rises to 9.4\,K for the powder samples.
The inset of Fig. \ref{fig:SF-T}(b) depicts an example of the definition of the onset $\Tc$.
The SF at 5\,K for sample 1a in the ZFC condition is 12\%. and the SF at 5\,K for samples 2a--4a  (20--250 \,$\mu$m) is enhanced and is over 80\% in the ZFC condition.
For the smallest sample 5a (0--20 \,$\mu$m), however, the SF is suppressed to 14\% at 5\,K.
On the other hand, the SF in the FC condition 
for sample 1a (single crystal) is only about 1\%.
The SF in the FC condition increases with decreasing sample size except for sample 5a.
For sample  5a, the SF in the FC condition is only 1\%.
The SF measured in the FC condition is correlated with the superconducting volume fraction (SVF).
Accordingly, we use the SF in the FC condition as a measure of the SVF.
The sample-size dependence of the SF measured in the FC condition 
 is summarized in Table \ref{tab:susceptibility}.
The results indicate that the O$_{2}$-annealing effects for the Fe$_{1+y}$Te$_{1-x}$S$_{x}$ 
strongly depends on the sample size, 
and there appears to be an optimum sample size 
to maximize the SVF.
  
\begin{table}[h]
\begin{center}
\caption{Shielding fraction at 5\,K in the FC condition (SF-FC) and onset $\Tca$ and onset $\Tcb$ defined in Fig. \ref{fig:SF-T}.
}
\begin{tabular}{clccc}
Sample & size & SF-FC (\%) & onset $\Tca$ (K) & onset $\Tcb$ (K) \\ \hline
1a & 3~3~0.5 mm$^3$ & 1.1 & 8.8 & 7.0 \\
2a &  106--250 $\mu$m & 3.2 & 9.4 & 8.4 \\
3a & 75--106 $\mu$m & 6.2 & 9.4 & 8.6 \\
4a & 20--75 $\mu$m & 11.0 & 9.4 & 8.8 \\
5a & 0--20 $\mu$m & 1.2 & 9.2 & 8.9 \\\end{tabular}
\label{tab:susceptibility}
\end{center}
\end{table}

\begin{figure}[b]
\begin{center}
\includegraphics[width=70mm]{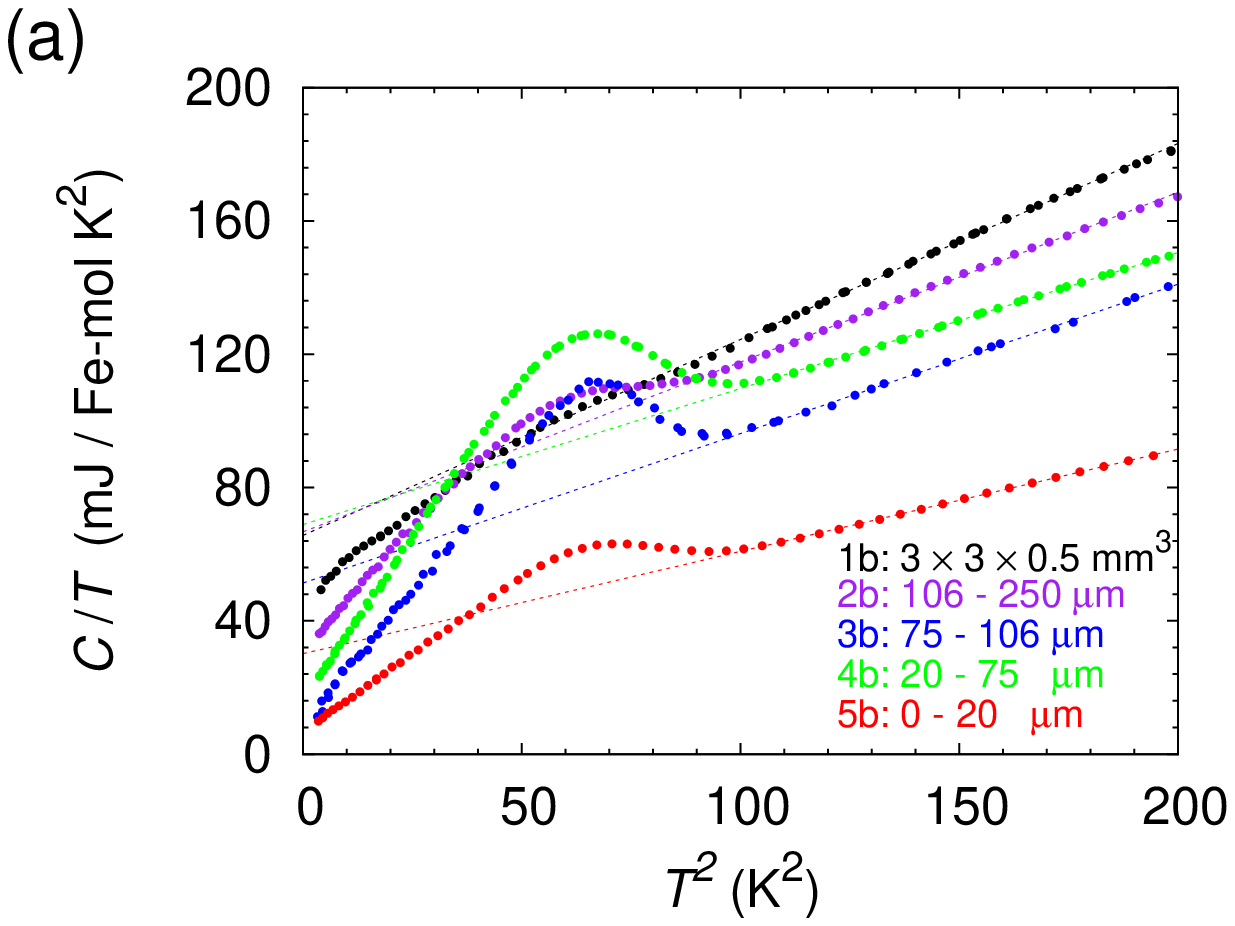}
\includegraphics[width=70mm]{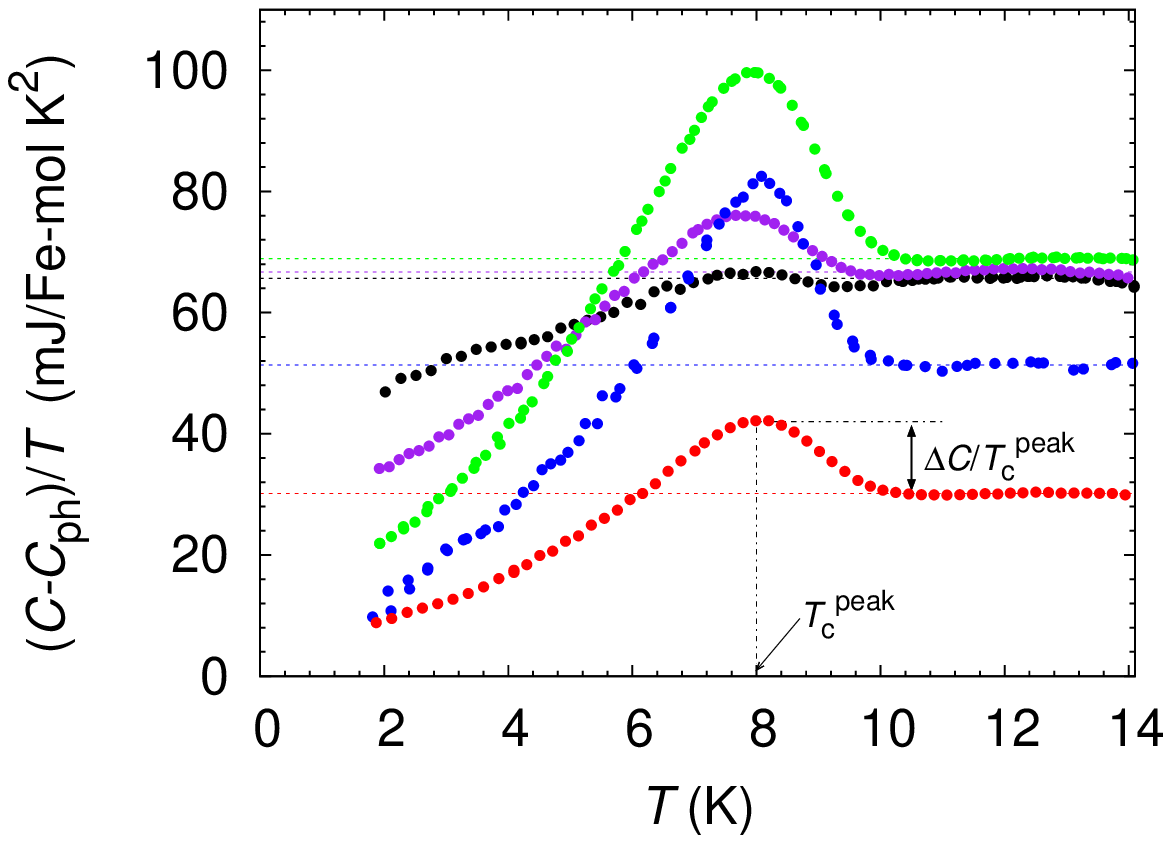}
\caption{(Color online) (a) $C/T$ of O$_{2}$-annealed Fe$_{1.01}$Te$_{0.91}$S$_{0.09}$ against $T^{2}$ plot. (b) $(C-\Cph)$ vs $T$ plot of Fe$_{1.01}$Te$_{0.91}$S$_{0.09}$.
 }
\label{fig:C-T}
\end{center}
\end{figure}

The specific heat provides quantitative and reliable information about the SVF.
The specific heat divided by temperature, $C/T$, against $T^2$ is 
plotted in Fig. \ref{fig:C-T} (a).
Peaks associated with the superconducting transition were 
observed in the O$_{2}$-annealed powder samples 2a - 5a, 
although such a peak was hardly observed in the single crystalline sample 1b.
As will be described later, these samples used include 
several kinds of impurities, e. g. FeTe$_{2}$.
It is, therefore, difficult to distinguish the specific heat of Fe$_{1.01}$Te$_{0.91}$S$_{0.09}$ 
from those of the impurities.
Here we analyze the specific heat data assuming the following simple formula 
for the temperature range of 100\,K$^2$ $\leq T^2 \leq$ 196\,K$^2$ ,
\begin{equation}
\frac{C}{T} = \gamma + \beta  T^2, 
\end{equation}
where $\gamma$  and  $\beta$ mainly come from
the electronic and the lattice specific heat coefficients of Fe$_{1.01}$Te$_{0.91}$S$_{0.09}$
, respectively.
Figure. \ref{fig:C-T}(b) shows the electronic part of the specific heat 
obtained from the data on the assumption of the functional form of equation (1).
This analysis apparently fails to extract the electronic contribution of Fe$_{1.01}$Te$_{0.91}$S$_{0.09}$
  because the entropy balance is not conserved in Fig. \ref{fig:C-T}(b).
Nevertheless, the peaks observed in the specific heat are so clear that their attributions to the superconducting transition appear to be plausible.
The onset temperature of the peak (onset $\Tc$), the peak temperature ($\Tc^{\mbox{\scriptsize peak}}$), and the parameters obtained from the fitting procedure are tabulated in Table \ref{tab:parameters}.
Note that fitting the experimental data to equation (1) 
yields $\gamma$ and $\beta$ including the contributions from the impurities.
The Debye temperature $\theta_{\mbox{\scriptsize D}}$ estimated from $\beta$ in sample 1b is 149 K, which is close to a Debye temperature of 141 K for the parent compound Fe$_{1.05}$Te reported in Ref.\citen{chen2009electronic}.
The onset temperature of the peak increases to about 10.2\,K with decreasing sample size for samples 2b--4b
 and that of sample 5b slightly decreases.
The peak temperature is about 8.0\,K in the sample 2b--5b.
The peak heights $\Delta C/\Tc^{\mbox{\scriptsize peak}}$ of samples 3b and 4b are 
  very close (31 mJ/Fe-mol K$^2$) and are the greatest among the samples.

\begin{table}[h]

\begin{center}
\caption{Parameters estimated by the specific heat measurements}
\begin{tabular}{clcccccc}
Sample & size & $\beta$ (mJ/K$^4$mol) & $\theta_{\mbox{\scriptsize D}}$ (K) & $\gamma$ {\scriptsize (mJ/Fe-mol K$^2$) }& onset-$\Tc$ (K) & $\Tc^{\mbox{\scriptsize peak}}$ (K) & $\Delta C/\Tc^{\mbox{\scriptsize peak}}$ {\scriptsize (mJ/Fe-mol K$^2$)} \\ \hline
1b & 3~3~0.5 mm$^3$ & 0.59 & 149 & 66 & (9.0) & (8.0) & - \\
2b & 106 - 250 $\mu$m & 0.51 & 156 & 67 & 9.5 & 7.8 & 9.3 \\
3b & 75 - 106 $\mu$m & 0.45 & 163 & 51 & 9.6 & 8.1 & 31 \\
4b & 20 -   75 $\mu$m & 0.41 & 169 & 69 & 10.2 & 8.0 & 31 \\
5b & 0 -   20 $\mu$m & 0.31 & 185 & 30 & 10.0 & 8.0 & 12 \\
\end{tabular}
\label{tab:parameters}
\end{center}
\end{table}

\begin{figure}[h]
\begin{center}
\includegraphics[width=\linewidth]{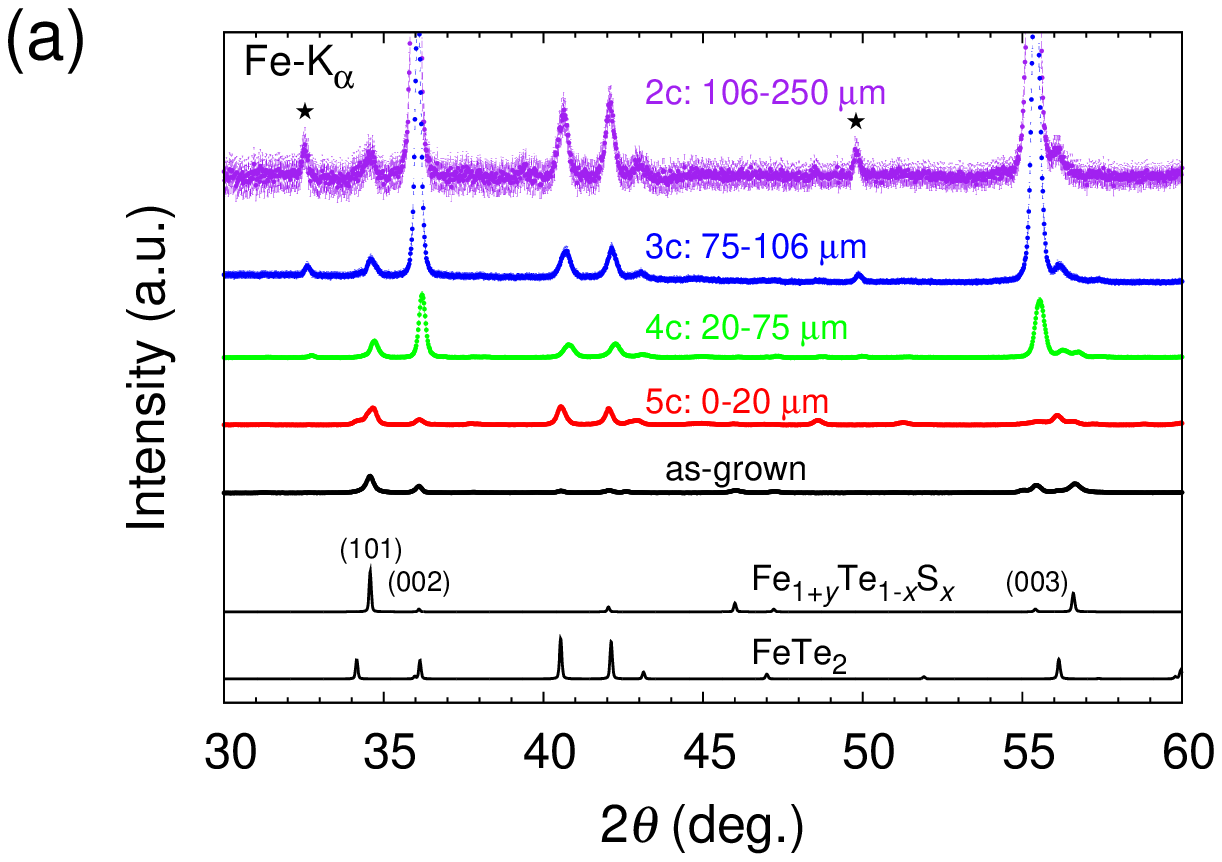}
\includegraphics[width=\linewidth]{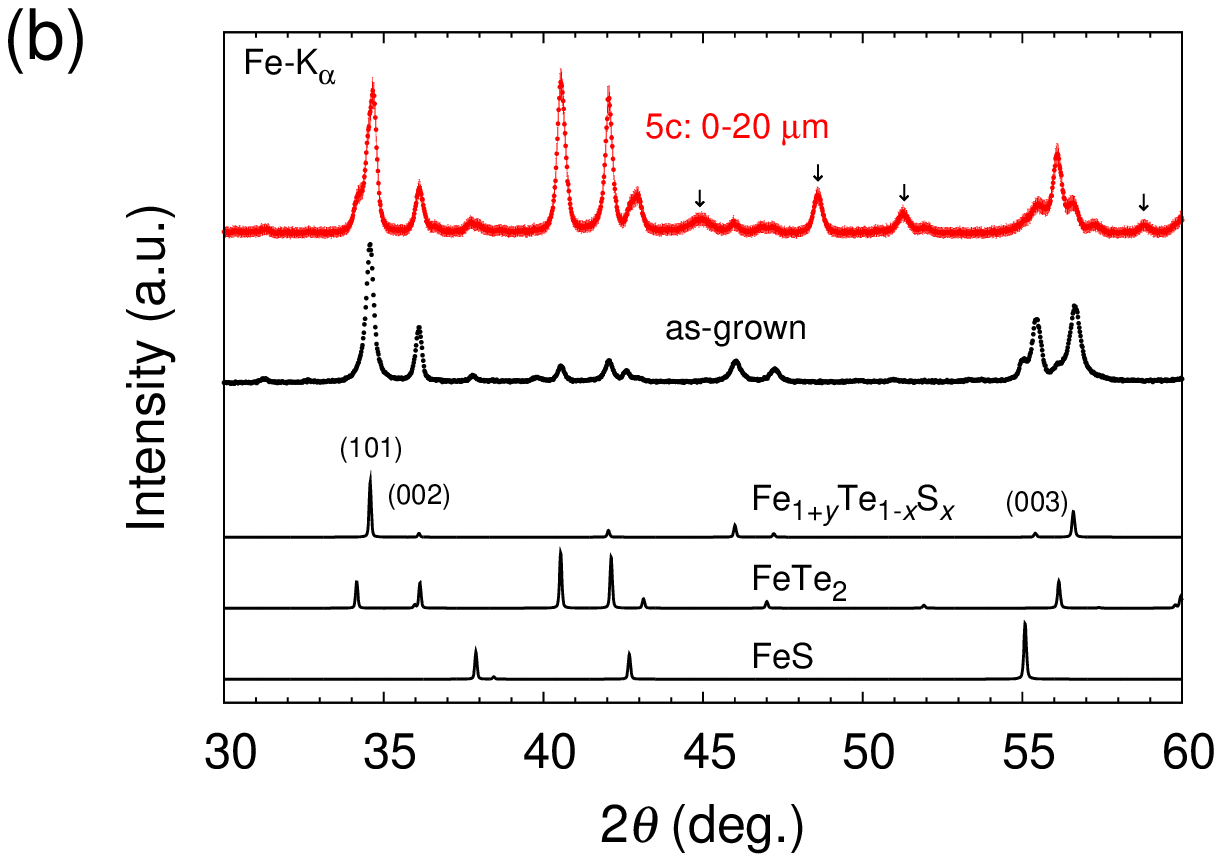}
\caption{(Color online) 
(a) X-ray powder diffraction patterns of size-classified O$_2$-annealed Fe$_{1.01}$Te$_{0.91}$S$_{0.09}$ and the as-grown sample.
(b) Enlarged view of data for the smallest sample 5c and the as-grown sample.
The data are scaled by the peak intensity of the (101) reflection.
Calculated spectra for Fe$_{1-x}$Te$_{1-x}$S$_{x}$, FeTe$_2$ ,and FeS are attached.
The cross marks represent the (00$l$) reflections of Fe-K$_{\beta}$ radiation.
The arrows represent peaks of other materials that could not be identified.
 }
\label{fig:XRD}
\end{center}
\end{figure}

The size dependence observed in the susceptibility and the specific heat measurements indicates that the O$_{2}$ annealing affects due to the distance from surface.
To identify the materials created in the samples through O$_{2}$ annealing, X-ray diffraction measurements were carried out.
Figure \ref{fig:XRD}(a) shows the X-ray diffraction patterns of size-classified O$_2$-annealed samples of Fe$_{1.01}$Te$_{0.91}$S$_{0.09}$, and those of an as-grown sample for comparison.
While the as-grown sample was ground for the measurements, the O$_2$-annealed samples were mounted on the glass plates without being ground.
Consequently, the X-ray diffraction data for O$_2$-annealed samples reflect the composition of the materials near the surface;
 the penetration depth of the X-ray used is about 5 $\mu$m.
The data were scaled by the (101) peak intensity because of quite a high intensity of $(00l)$ peaks owing to the O$_{2}$-annealed samples being naturally oriented.
Figure \ref{fig:XRD}(b) shows an enlarged view of the data for the as-grown sample and the smallest sample 5c.
In the as-grown sample, small peaks from FeS and FeTe$_2$ are also seen in addition to the peaks of {Fe$_{1.01}$Te$_{0.91}$S$_{0.09}$}.
Since sulfur has a solubility limit\cite{mizuguchi2011single}, the surplus sulfur might react and form these impurities.
Therefore,  these impurities could be inevitable in the single crystalline sample grown by the self flux method.
Besides this, the two peaks at 40.5$^{\circ}$ and 42.1$^{\circ}$ attributable to FeTe$_{2}$ are clearly observed for the O$_2$-annealed samples.
 FeTe$_{2}$ is probably generated by the O$_{2}$ annealing at surfaces of Fe$_{1.01}$Te$_{0.91}$S$_{0.09}$.
FeTe$_2$ has been observed also in Fe$_{1+y}$Te$_{1-x}$S$_{x}$ annealed in the oxygen at temperatures higher than 300 $^\circ$C\cite{mizuguchi2010evolution}.
For all of the samples annealed in oxygen atmosphere, reflections attributable to Fe$_{2}$O$_{3}$ or Fe$_{3}$O$_{4}$ were not observed.
In the smallest powder sample 5c, many additional peaks were observed.
Although they have not been identified, other impurity peaks indicated by arrows are also seen.
Other unidentified materials are produced by O$_2$ annealing especially in the smallest sample 5c. 

\section{Discussion}
\begin{figure}[h]
\begin{center}
\includegraphics[width=\linewidth]{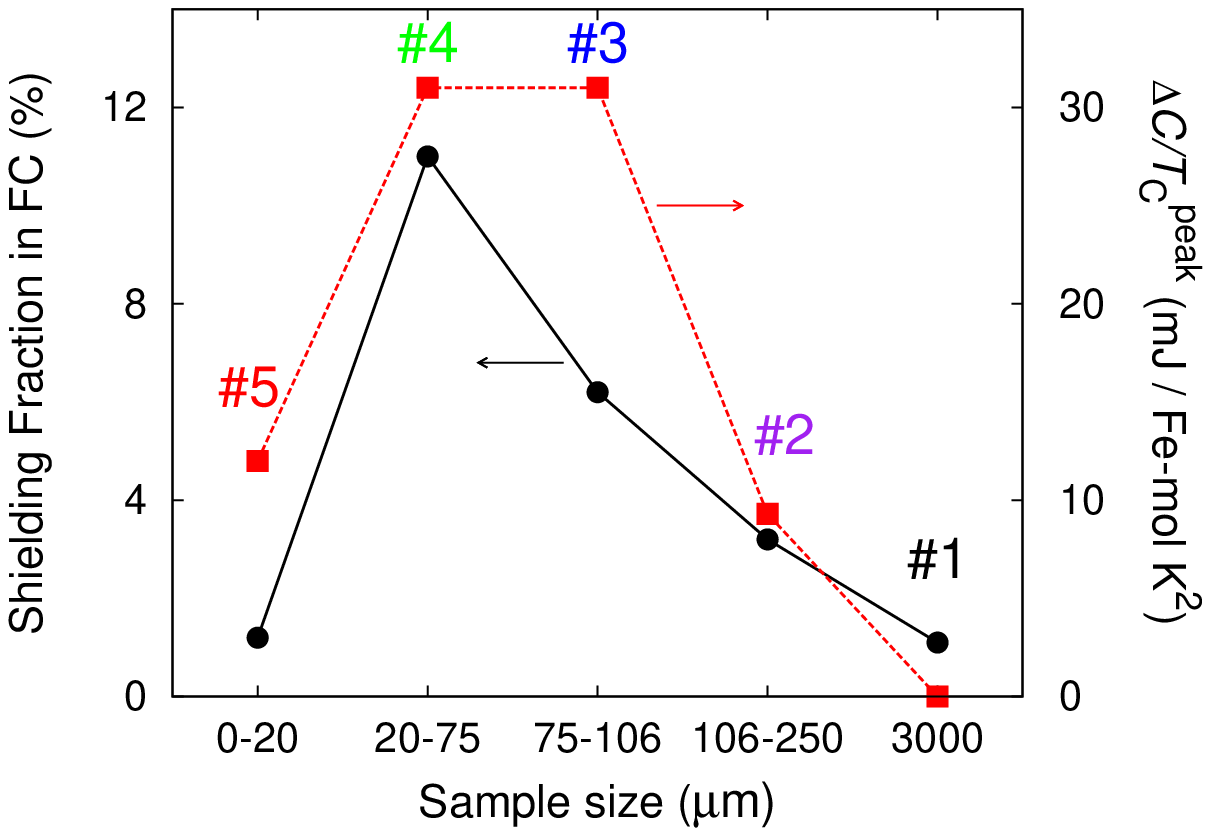}
\caption{(Color online) 
Sample size dependences of shielding fraction (SF) measured in FC condition and $\Delta C/\Tc^{\mbox{\scriptsize peak}}$ evaluated from specific heat measurements.
 }
\label{fig:SVF}	
\end{center}
\end{figure}

The sample-size dependences of the SVF in O$_2$-annealed Fe$_{1.01}$Te$_{0.91}$S$_{0.09}$ 
 evaluated from FC susceptibility and from specific heat measurements show almost the same trend as illustrated in Fig. \ref{fig:SVF}.
The SVF becomes greater with decreasing sample size down to 20 $\mu$m,
but it becomes lower in smaller samples. 
The results indicate two types of effects of O$_2$ annealing in Fe$_{1.01}$Te$_{0.91}$S$_{0.09}$:
 creation and deterioration of superconducting regions.
It is unable to evaluate the accurate SVF from these results because of the difficulty of the analysis for the specific heat as mentioned above.
The  volume fraction, however, may be roughly estimated to be dozens of percent of order
 by assuming the BCS theory, comparing
  $\Delta C/\Tc^{\mbox{\scriptsize peak}}$ observed in sample 3b and 4b are from 0.4$\gamma$--0.6$\gamma$, which is $1.43\gamma$ in the BCS theory.

\begin{figure}[h]
\begin{center}
\includegraphics[width=\linewidth]{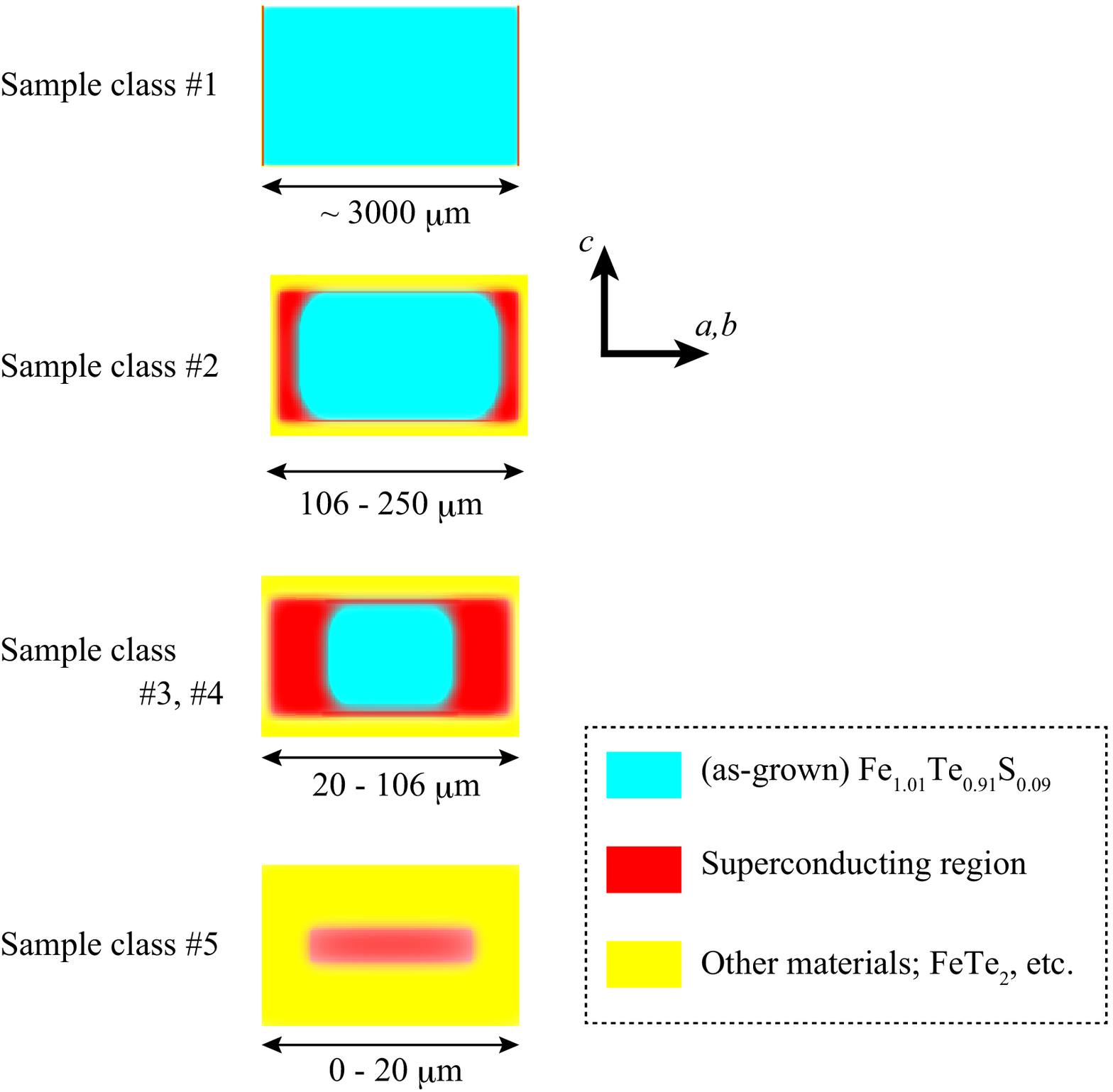}
\caption{(Color online) 
Schematic  sectioned drawings of O$_{2}$-annealed Fe$_{1+y}$Te$_{1-x}$S$_{x}$.
Assuming that the thickness of the superconducting region and that of the other produced materials region are fixed, the trend of the sample-size dependence of the superconducting volume fraction (SVF) can be explained.
 }
\label{fig:drawing}	
\end{center}
\end{figure}

From the sample-size dependence of the superconducting volume fraction, the superconducting regions probably exist within a thickness of a few tens of micro meters from the surface.
Figure \ref{fig:drawing} shows schematic diagrams of the O{$_2$}-annealing effects we propose.
In order to construct these diagrams, we have made the following assumptions:
\begin{enumerate}
\item O$_2$ annealing mostly affects regions close to sample surfaces.
\item As a consequence of the O$_2$ annealing, two distinct layers are formed near the surfaces.
The outer layer contains a large amount of non-superconducting materials such as FeTe$_2$ while the inner layer mostly consists of superconducting regions that encloses almost the whole sample. 
The core region surrounded by the two layers remains unchanged after the annealing.
\item The thicknesses of those two layers are
anisotropic due to the anisotropy of the crystal structure, but are
 independent of the sample size, and may be regarded as somewhat similar to a penetration depth of the annealing effects.
\end{enumerate}
In fact, the formation of layers responsible for superconductivity,which enclose almost the whole sample, 
 is consistent with the large difference between the SF in ZFC and FC conditions.
Indeed, these assumptions reasonably well account for the observations concerning the SVF.
The SVF will increase with decreasing sample size when the dimensions of the sample is considerably larger than the thickness of the two layers. 
On the other hand, when the dimensions of the sample becomes less than the thickness of the layers, the inner superconducting layer will naturally diminish and the SVF will decrease.
As shown in Table \ref{tab:parameters}, the electronic specific heat coefficient $\gamma$ of the sample class \#5 is relatively small.
This result is consistent with our view shown in Fig \ref{fig:drawing} because FeTe$_2$, which is an insulator ($\gamma \sim 0$), is one of the major impurities in the smallest sample class \#5\cite{westrum1959heat}.
As stated earlier, the X-ray diffraction measurements did not detect oxidized iron such as Fe{$_2$O$_3$} or Fe{$_3$O$_4$} in the O$_2$-annealed samples.
Nevertheless, we observed an oxygen component at surfaces of the O$_2$-annealed samples in EPMA measurements.
We conjecture that oxygen could be diffused in Fe$_{1+y}$Te$_{1-x}$S$_{x}$ and intercalated between adjacent atomic layers  as suggested in Ref.\citen{mizuguchi2010evolution}.
Such intercalation perhaps reduces the effects of excess iron and induces superconductivity.

\section{Conclusions}
We have investigated the sample-size dependence of the O$_{2}$-annealing effects
  by means of ZFC and the FC susceptibility and specific heat measurements.
 We observed a jump of the specific heat around superconducting transition temperature $\Tc$ in the O$_2$-annealed small samples.
As far as we know, this is the first observation of the bulk superconductivity in Fe$_{1+y}$Te$_{1-x}$S$_{x}$.
 It implies that the superconducting region induced by O$_2$ annealing is distributed probably near the surface of samples over a length of a few tens of micro meters in the annealing condition of 1\,atm, 2\,hours, and 200\,$^\circ$C.

\section*{Acknowledgements}
This work was partly carried out under the Visiting Researcher's Program of the Institute for Solid State Physics, the University of Tokyo.
We would like to thank H. Yoshizawa for his support in the specific heat measurements.

We are grateful to K. Motoya and T. Moyoshi for allowing us to use electron probe microanalyzer (EPMA) and equipment for sample synthesis.



\begin{thebibliography}{10}

\bibitem{kamihara2006iron}
Y.~Kamihara, H.~Hiramatsu, M.~Hirano, R.~Kawamura, H.~Yanagi, T.~Kamiya, and
  H.~Hosono: J. Am. Chem. Soc. {\bfseries 128} (2006) 10012.

\bibitem{kamihara2008iron}
Y.~Kamihara, T.~Watanabe, M.~Hirano, and H.~Hosono: J. Am. Chem. Soc.
  {\bfseries 130} (2008) 3296.

\bibitem{deguchi2012physics}
K.~Deguchi, Y.~Takano, and Y.~Mizuguchi: Sci. Technol. Adv. Mater. {\bfseries
  13} (2012).

\bibitem{li2009first}
S.~Li, C.~de~La~Cruz, Q.~Huang, Y.~Chen, J.~Lynn, J.~Hu, Y.-L. Huang, F.-C.
  Hsu, K.-W. Yeh, M.-K. Wu, et~al.: Phys. Rev. B {\bfseries 79} (2009) 054503.

\bibitem{bao2009tunable}
W.~Bao, Y.~Qiu, Q.~Huang, M.~Green, P.~Zajdel, M.~Fitzsimmons, M.~Zhernenkov,
  S.~Chang, M.~Fang, B.~Qian, et~al.: Phys. Rev. Lett. {\bfseries 102} (2009)
  247001.

\bibitem{hu2009superconductivity}
R.~Hu, E.~S. Bozin, J.~Warren, and C.~Petrovic: Phys. Rev. B {\bfseries 80}
  (2009) 214514.

\bibitem{zajdel2010phase}
P.~Zajdel, P.-Y. Hsieh, E.~E. Rodriguez, N.~P. Butch, J.~D. Magill,
  J.~Paglione, P.~Zavalij, M.~R. Suchomel, and M.~A. Green: J. Am. Chem. Soc.
  {\bfseries 132} (2010) 13000.

\bibitem{koz2013low}
C.~Koz, S.~R{\"o}{\ss}ler, A.~A. Tsirlin, S.~Wirth, and U.~Schwarz: Phys. Rev.
  B {\bfseries 88} (2013) 094509.

\bibitem{mizuguchi2011single}
Y.~Mizuguchi, K.~Deguchi, T.~Ozaki, M.~Nagao, S.~Tsuda, T.~Yamaguchi, and
  Y.~Takano: Appl. Supercond. {\bfseries 21} (2011) 2866.

\bibitem{mizuguchi2010moisture}
Y.~Mizuguchi, K.~Deguchi, S.~Tsuda, T.~Yamaguchi, and Y.~Takano: Phys. Rev. B
  {\bfseries 81} (2010) 214510.

\bibitem{mizuguchi2010evolution}
Y.~Mizuguchi, K.~Deguchi, S.~Tsuda, T.~Yamaguchi, and Y.~Takano: Europhys.
  Lett. {\bfseries 90} (2010) 57002.

\bibitem{kawasaki2011pressure}
Y.~Kawasaki, Y.~Mizuguchi, K.~Deguchi, T.~Watanabe, T.~Ozaki, S.~Tsuda,
  T.~Yamaguchi, and Y.~Takano: Physica C {\bfseries 471} (2011) 611.

\bibitem{awana2011superconductivity}
V.~Awana, A.~Pal, A.~Vajpayee, B.~Gahtori, and H.~Kishan: Physica C {\bfseries
  471} (2011) 77.

\bibitem{deguchi2014excess}
K.~Deguchi, A.~Yamashita, T.~Yamaki, H.~Hara, S.~Demura, S.~Denholme,
  M.~Fujioka, H.~Okazaki, H.~Takeya, T.~Yamaguchi, et~al.: Journal of Applied
  Physics {\bfseries 115} (2014) 053909.

\bibitem{deguchi2011alcoholic}
K.~Deguchi, Y.~Mizuguchi, Y.~Kawasaki, T.~Ozaki, S.~Tsuda, T.~Yamaguchi, and
  Y.~Takano: Supercond. Sci. Technol. {\bfseries 24} (2011) 055008.

\bibitem{deguchi2012clarification}
K.~Deguchi, D.~Sato, M.~Sugimoto, H.~Hara, Y.~Kawasaki, S.~Demura, T.~Watanabe,
  S.~Denholme, H.~Okazaki, T.~Ozaki, et~al.: Supercond. Sci. Technol.
  {\bfseries 25} (2012) 084025.

\bibitem{noji2012specific}
T.~Noji, M.~Imaizumi, T.~Suzuki, T.~Adachi, M.~Kato, and Y.~Koike: J. Phys.
  Soc. Jpn. {\bfseries 81} (2012) 054708.

\bibitem{imaizumi2012superconductivity}
M.~Imaizumi, T.~Noji, T.~Adachi, and Y.~Koike: J. Phys.: Conf. Ser. {\bfseries
  400} (2012) 022034.

\bibitem{kawasaki2012phase}
Y.~Kawasaki, K.~Deguchi, S.~Demura, T.~Watanabe, H.~Okazaki, T.~Ozaki,
  T.~Yamaguchi, H.~Takeya, and Y.~Takano: Solid State Commun. {\bfseries 152}
  (2012) 1135.

\bibitem{sun2012effects}
Y.~Sun, T.~Taen, Y.~Tsuchiya, Z.~Shi, and T.~Tamegai: Supercond. Sci. Technol.
  {\bfseries 26} (2012) 015015.

\bibitem{rodriguez2011chemical}
E.~E. Rodriguez, C.~Stock, P.-Y. Hsieh, N.~P. Butch, J.~Paglione, and M.~A.
  Green: Chemical Science {\bfseries 2} (2011) 1782.

\bibitem{shi2010accurate}
Q.~Shi, C.~L. Snow, J.~Boerio-Goates, and B.~F. Woodfield: J. Chem. Therm.
  {\bfseries 42} (2010) 1107.

\bibitem{martin1960specific}
D.~L. Martin: Can. J. Phys. {\bfseries 38} (1960) 17.

\bibitem{chen2009electronic}
G.~Chen, Z.~Chen, J.~Dong, W.~Hu, G.~Li, X.~Zhang, P.~Zheng, J.~Luo, and
  N.~Wang: Phys. Rev. B {\bfseries 79} (2009) 140509.

\bibitem{westrum1959heat}
E.~Westrum~Jr, C.~Chou, and F.~Gronvold: J. Chem. Phys. {\bfseries 30} (1959)
  761.

\end{thebibliography}
\end{document}